\documentclass[a4paper, amsfonts, amssymb, amsmath, reprint, showkeys, nofootinbib, twoside]{revtex4-1}

\pdfoutput=1

\usepackage[compat=1.0.0]{tikz-feynman}
\usepackage{subfigure}

\usepackage[english]{babel}
\usepackage[utf8]{inputenc}
\usepackage[colorinlistoftodos, color=green!40, prependcaption]{todonotes}
\usepackage[pdftex, pdftitle={Article}, pdfauthor={Author}]{hyperref} 
\begin{document}
\title{Cosmogenic neutrino fluxes under the effect of active-sterile secret interactions}

\author{Damiano Fiorillo, Gennaro Miele, Stefano Morisi}
    \affiliation{Dipartimento di Fisica {\it "Ettore Pancini"}, Universit\`a degli studi di Napoli Federico II, Complesso Univ. Monte S. Angelo, I-80126 Napoli, Italy}
    \affiliation{INFN - Sezione di Napoli, Complesso Univ. Monte S. Angelo, I-80126 Napoli, Italy}
    
\author{Ninetta Saviano}
	\affiliation{INFN - Sezione di Napoli, Complesso Univ. Monte S. Angelo, I-80126 Napoli, Italy}

\date{\today} 

\begin{abstract}
Ultra High Energy cosmogenic neutrinos may represent a unique opportunity to unveil possible new physics interactions once restricted to the neutrino sector only. In the present paper we study the observable effects of a secret active-sterile interactions, mediated by a pseudoscalar, on the expected flux of cosmogenic neutrinos. The results show that for masses of sterile neutrinos and pseudoscalars of hundreds MeV, necessary to evade cosmological, astrophysical and elementary particle constraints, the presence of such new interactions can significantly change the energy spectrum of cosmogenic neutrinos at Earth in the energy range from PeV to ZeV. Interestingly, the distortion of the spectrum results to be detectable at apparatus like GRAND or ARIANNA if the mediator mass is around 250 MeV and the UHECRs are dominated by the proton component. Larger mediator masses or a chemical composition of UHECRs dominated by heavier nuclei would require much larger cosmic rays apparatus which might be available in future.
\end{abstract}

\maketitle

\section{Introduction}
The neutrino sector still represents a partially unknown territory. Fundamental questions like the nature of neutrinos (Dirac or Majorana) or the possible connection of their small masses with physics Beyond the Standard Model (BSM) can represent possible windows on new physics. 
In the last decade increasing attention has been devoted to high energy astrophysical neutrinos, after the observation of the first events at the IceCube detector \cite{Aartsen:2013jdh}. These astrophysical fluxes are of extreme relevance for the neutrino sector since they provide a powerful tool of investigation for Beyond Standard Model physics, such as sterile neutrinos, Lorentz violations and non standard model interactions.  In this context, cosmogenic neutrinos, that are mainly expected to have energies  up to $10^{12}\,$GeV, could represent a unique opportunity. 

Cosmogenic neutrinos are produced by the photo-hadronic interactions of Ultra High Energy Cosmic rays  (UHECRs), whose  precise chemical composition and origin is still unknown, with the Cosmic Microwave Background (CMB) \cite{Beresinsky:1969qj}. The same process is also responsible for a depletion in the flux of UHECRs, which is known in literature as the GZK cutoff \cite{Greisen:1966jv, Zatsepin:1966jv}. For this reason cosmogenic neutrinos are also known as GZK-neutrinos and they have been extensely studied in a number of works, see for instance \cite{Berezinsky:1998ft,Nagano:2000ve,Engel:2001hd,Kusenko:2001gj,Anchordoqui:2002hs,Fodor:2002hy,Kalashev:2002kx,Semikoz:2003wv,Fodor:2003ph,Ave:2004uj,Seckel:2005cm,DeMarco:2005kt,Allard:2006mv,Becker:2007sv,Anchordoqui:2007fi,Berezinsky:2010xa,Ahlers:2010fw,Katz:2011ke,Gelmini:2011kg,Ng:2014pca,Murase:2014tsa,Heinze:2015hhp,Aloisio:2015ega,Halzen:2016gng,Cherry:2018rxj,Vitagliano:2019yzm}.

The desirable observation of high-energy cosmogenic neutrinos would be particularly relevant in order to determine the origin of UHECRs. Several cosmic rays apparatus, like HiRes \cite{Abbasi:2007sv} and the Pierre Auger Observatory (PAO) \cite{Abraham:2004dt, Abraham:2008ru} for example, have already tried to perform such a measurement, and others are planned to do it in future with much better chances, like for instance  GRAND \cite{Alvarez-Muniz:2018bhp}, ARIANNA \cite{Anker:2019rzo}, ARA \cite{Allison:2015eky} and JEM-EUSO \cite{Adams:2013vea}.  Unfortunately, a possible additional difficulty lies in the fact that cosmogenic neutrino fluxes have a strong dependence on the chemical composition of cosmic rays, as shown for instance in \cite{Ahlers:2012rz}. This is particularly important since recent results suggest that the chemical composition is actually mixed, containing significant amounts of heavier nuclei  rather than simply protons \cite{Castellina:2019huz}, and unfortunately for heavier nuclei one expects a suppression in the neutrino production. For this reason, in this work we have analyzed two benchmark scenarios for the chemical composition of the UHECRs, namely the cases in which the dominant component is either protons or Helium nuclei. 

In view of the above considerations a future detection of cosmogenic neutrinos would allow us to infer crucial astrophysical informations concerning  UHECRs chemical composition,  interactions and origin. However, at the same time the shape of the cosmogenic neutrino spectrum, that could be strongly distorted by the presence of new physics, could unveil the presence of these new interactions at least in the neutrino sector. In particular, any kind of BSM interactions  involving  active neutrinos  $\nu$ could modify the expected spectrum and rate of cosmogenic neutrinos at Earth.  Examples of such new couplings discussed in literature are:\\
\noindent i) \textit{Non Standard Interactions} (NSI) for active neutrinos $\nu\nu \to ff$ \cite{Davidson:2003ha,Antusch:2008tz,Miranda:2004nb,Fornengo:2001pm,Huber:2001zw,Barranco:2005ps,Farzan:2017xzy,Ribeiro:2007ud,Coloma:2015kiu,deGouvea:2015ndi,Forero:2011pc,Mangano:2006ar},  where $f$ denotes quarks or charged leptons;\\
\noindent ii)\textit{Secret interactions} (SI) mediated by some new boson (scalar or vector), and just restricted to the active neutrino sector $\nu\nu \to \nu\nu$ \cite{Ng:2014pca,Ioka:2014kca,Bakhti:2018avv,Kolb:1987qy, Archidiacono:2013dua,Laha:2013xua,Forastieri:2019cuf,Bustamante:2020mep,Blum:2014ewa,Murase:2019xqi,Babu:2019iml};\\
\noindent iii)Secret interactions just restricted to the sterile neutrino sector $\nu_s\nu_s \to \nu_s\nu_s$ for different sterile neutrino mass scales, \cite{Hannestad:2013ana,Dasgupta:2013zpn,Archidiacono:2014nda,Saviano:2014esa,Mirizzi:2014ama,Cherry:2014xra,Chu:2015ipa,Archidiacono:2016kkh,Cherry:2016jol,Forastieri:2017oma,Chu:2018gxk,Jeong:2018yts,deGouvea:2019phk}. \\
\noindent iv) Secret interactions involving active and sterile neutrinos simultaneously $\nu\nu \to \nu_s\nu_s$\cite{Babu:1991at,Shoemaker:2015qul}\\

In this paper we consider  a scheme of SI similar to the point iv, wherein the new interaction, mediated  by a new pseudoscalar boson, intervene both active and sterile neutrinos. While \cite{Babu:1991at} explores the effects of this interaction on primordial nucleosynthesis, and \cite{Shoemaker:2015qul} studies the effects of the interaction on neutrinos in the IceCube energy range of interest, we have analyzed the effects of this interaction on cosmogenic neutrino fluxes to question their observability. We will assume throughout that both the active and the sterile neutrinos are Majorana particles: this is also the reason why we have to choose a pseudoscalar mediator. In fact the scalar contraction $\bar{\nu} \nu_s$ is antihermitean and therefore not admissible as a possible interaction operator: therefore the only possible contraction is $\bar{\nu} \gamma_5 \nu_s$. To preserve parity, we take the mediator to be a pseudoscalar. In particular we study  the distortion implied by such new coupling on the expected cosmogenic neutrino flux estimating the possibility to measure this effect in apparatus like GRAND \cite{Alvarez-Muniz:2018bhp}. For the sake of simplicity we will assume just one sterile neutrino (hereafter denoted by $\nu_s$) coupling with the active sector {\it via} this new interaction. Moreover, with the aim to catch the main implications of such scenario on cosmogenic neutrino flux we consider an active neutrino only (hereafter denoted by $\nu$), hence focussing on a more simple $1+1$ framework.  However it would be straightforward to extend our analysis to three active neutrinos even though we do not expect that the main results, here  obtained, would be drastically changed in a more  realistic $3+1$ framework. The interaction term then becomes
\begin{eqnarray}
\mathcal{L}_{\rm{SI}} = \lambda \, \overline{\nu} \gamma_5 \nu_s \varphi \,,\label{coupling}
\end{eqnarray}
where $\lambda$ is a dimensionless free coupling.
\vskip5.mm

\section{Cosmogenic neutrino flux at Earth without secret interactions}

As described above, cosmogenic neutrinos are produced by the scattering of high energy protons from the cosmic rays with the CMB photons. The cosmogenic neutrino flux $\phi_\nu$, expected to be isotropic, can  be parameterized in the form
\begin{equation} \label{cosmpar}
\frac{d\phi_\nu}{dEd\Omega}= \int\frac{dz'}{H(z')} F\left[z', E(1+z')\right]\,,
\end{equation}
where $F\left[z', E(1+z')\right]$ is the number of neutrinos produced per unit time per unit energy interval per unit solid angle per unit volume at redshift $z'$ and with comoving energy $E(1+z')$. We use as a reference the spectrum proposed in \cite{Ahlers:2012rz}, which constitutes a lower bound for the cosmogenic neutrino spectrum. This is a conservative hypothesis, since a higher flux would make easier to detect the effects of the interaction.

The quantity $F\left[z',E (1+z')\right]$ depends of course on the proton spectrum, which is itself the solution of a Boltzmann equation (see \cite{Ahlers:2012rz}), which takes into account the proton energy losses due to Bethe-Heitler processes and their depletion due to $p\gamma$ processes. This calculation has been performed for example in \cite{Ahlers:2012rz}, which provides the neutrino spectrum expected at Earth. It is important to notice that they assume a cosmic ray spectrum purely made of protons and emitted by sources whose density follows the star formation evolution. In other words, the proton luminosity can be written as $\mathcal{L}_p (z,E)=\mathcal{H} (z) Q_p (E)$, where $Q_p (E)$ is the proton injection spectrum from the single sources and $\mathcal{H} (z)$ is the Star Forming rate \cite{Hopkins:2006bw}
\begin{equation}\label{SFR}
\mathcal{H}(z) = \begin{cases}
(1+z)^{3.4} \;  \; \; \; \; \; \; \; \; \; \; \; \; \; \; \; \; \; \; z\leq1; \\ N_1 (1+z)^{-0.3} \; \; \; \; \; 1<z\leq4;\\ N_1 N_4 (1+z)^{-3.5} \; \; \; \; \; \; \; z>4, \end{cases}
\end{equation}
where $N_1=2^{3.7}$ and $N_4=5^{3.2}$.
The proton luminosity $\mathcal{L}_p(z,E)$ works as an input to the Boltzmann equation which provides the function $F\left[z',E (1+z')\right]$ in Eq. \eqref{cosmpar}. Since Ref. \cite{Ahlers:2012rz} only furnishes the final neutrino spectrum at Earth, and does not provide the proton spectrum $Q_p (E)$ at each redshift, in principle one would not be able to reproduce the function $F\left[z', E(1+z')\right]$ in Eq.\,\eqref{cosmpar}. However, at sufficiently high energies the mean free path for $p\gamma$ interaction becomes so small that neutrinos can be assumed to be produced exactly at the same place in which the emission of the protons occurs. If we make the further hypothesis that at each redshift the injection spectrum of the protons has exactly the same form, with a redshifted energy, the function $F\left[z',E(1+z')\right]$ takes on the form
\begin{equation}
\label{ansatz}
F[z',E(1+z')]=\rho(z') f[E(1+z')]\,,
\end{equation}
where $\rho(z')$ is proportional to the Star Forming Rate given in Eq.\,(\ref{SFR}). 

After this simplification, there is a unique function $f[E(1+z')]$ which reproduces the spectrum obtained by \cite{Ahlers:2012rz}. While the effects of these simplifications may be relevant at lower energies, they should be almost irrelevant in the high energy part of the cosmogenic spectrum, where the hypothesis of a small mean free path is natural. In particular, the mean free path for $p\gamma$ interactions is typically of the order of $50$ Mpc, which is small compared to the cosmological distances. 

The inversion of Eq.\,\eqref{cosmpar} under the ansatz of Eq.\,\eqref{ansatz} is in principle possible in an exact way through the use of a Mellin transform. In fact, the equation takes the form
\begin{equation}
\frac{d\phi_\nu}{dEd\Omega}= \int\frac{dz'}{H(z')} \rho(z') f[E(1+z')].
\end{equation}
However, due to the differential properties of the functions involved, this method is very hard to apply, because of the very fast oscillations of the Mellin transform. It is however possible to obtain a very good approximation by observing that the integral kernel connecting $f(E)$ to the observed spectrum, which is $\rho(z)/H(z)$, is a peaked function around $z\simeq 1$. Thus, under the assumption that $f[E(1+z)]$ depends on the redshift more slowly than $\rho(z)/H(z)$, we may take it out of the integral evaluating it at redshift $z=1$, finding
 \begin{equation} \label{newans}
 f(2E)= \frac{d\phi_\nu}{dEd\Omega} \frac{1}{\int\frac{dz'}{H(z')} \rho(z')}.
\end{equation}
We have numerically checked that this approximation gives good results by comparing the expected spectrum at Earth computed with Eq.\,\eqref{newans} with the input spectrum.

\section{Cross sections}

With the inclusion of secret interactions given in Eq.\,(\ref{coupling}), the cosmogenic neutrino spectra at Earth could change, depending on the free parameters of the new interaction, namely  the coupling $\lambda$, the masses  $M_\varphi$ of the scalar $\varphi$  and of the sterile neutrino $m_s$. In the following,  we provide  the cross sections for the processes considered in this work.

In the computation of the cross sections we should in principle consider initial and final states in the form of mass eigenstates. However, in the hypothesis that the active-sterile mixing angle $\theta_{as}$ is sufficiently small $\theta_{as}\ll 1$, the effects coming from taking this into account will be small corrections only, proportional at least to the square of the mixing angle $\theta_{as}^2$. The cross sections are therefore computed without any correction coming from mixing angles. Therefore the initial and final states may be taken directly as mass eigenstates.

\begin{figure} 
\begin{center}
\includegraphics[width=0.45\textwidth]{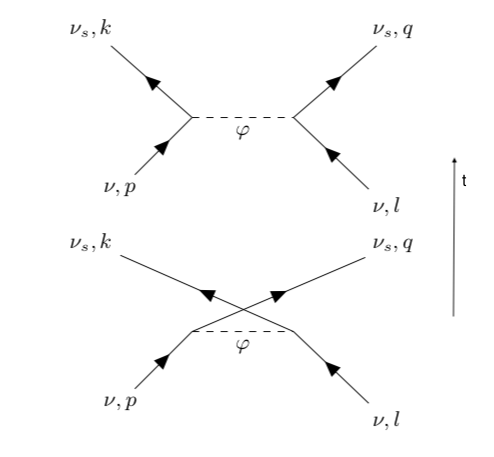}
\caption{Feynman diagrams for the scattering of two active neutrinos through the secret interaction.}\label{feyn1}
\end{center}
\end{figure}

\begin{figure} 
\begin{center}
\includegraphics[width=0.45\textwidth]{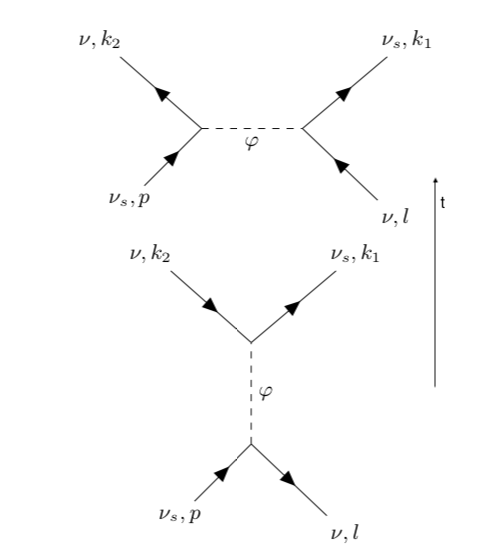}
\caption{Feynman diagrams for the scattering of an active and a sterile neutrino through the secret interaction.}\label{feyn2}
\end{center}
\end{figure}

 At tree level the new processes introduced by our new interaction are the four particle collisions $\nu+\nu\to \nu_s+\nu_s$, $\nu+\nu_s\to \nu+\nu_s$ and $\nu_s+\nu_s\to \nu+\nu$. Among these, the processes relevant for the experimental signatures we are looking for are the first two, shown in Fig.s\,\ref{feyn1} and\,\ref{feyn2} . 

In fact, since the Cosmic Neutrino Background (CNB) only involves active neutrinos, all collisions must involve at least one active neutrino in their initial state. As we will discuss in Section V, we will choose a range of parameters for which a sterile component in the CNB results to be negligible. The reason is that we wll choose sterile neutrinos so massive that their distribution becomes Boltzmann suppressed before the Big Bang Nucleosynthesis. The process $\nu+\nu_s\to \nu+\nu_s$ is still relevant, even though the cosmogenic neutrinos are active in flavor, because a sterile neutrino originated from mixing or from a previous collision of an active neutrino with the background might still in principle produce a relevant active flux. 

Considering the process $\nu+\nu\to \nu_s+\nu_s$ , the squared amplitude written in terms of the Mandelstam invariants $s=(p+l)^2$, $t=(p-k)^2$ and $u=(p-q)^2$ (see Figure \ref{feyn1}) is 
\begin{eqnarray} \label{amplit}
&&|\mathcal{M}_{aa\to ss}|^2 =\lambda^4 \left[ \frac{[t-(m-m_s)^2]^2}{(t-M_\varphi^2)^2+\Gamma^2 M_\varphi^2}
+ \frac{[u-(m-ms)^2]^2}{(u-M_\varphi^2)^2+\Gamma^2 M_\varphi^2} \right.\nonumber\\
&&- \frac{2[(t-M_\varphi^2)(u-M_\varphi^2)+\Gamma^2 M_\varphi^2]}{[(t-M_\varphi^2)^2+\Gamma^2 M_\varphi^2] [(u-M_\varphi^2)^2+\Gamma^2 M_\varphi^2]}  \nonumber\\
&&\times \left(\frac{(t-m^2-m_s^2)^2}{4} +\frac{(u-m^2-m_s^2)^2}{4}  \right.\nonumber\\
&&-\left. \left. \frac{s^2}{4}+s(m^2+m_s^2-m \, m_s) -2m^2 m_s^2 \right)  \right]
\end{eqnarray}
where $m$ is the mass of the active neutrino $\nu$ of CNB, $\Gamma$ is the decay rate of the scalar mediator given below, and $M_\varphi$ is its mass. We remind that the squared amplitude  depends on two Mandelstam invariants only, since the third is connected to the others by the relation $s+t+u=2(m^2+m_s^2)$. The total cross section for the $\nu+\nu\to \nu_s+\nu_s$ interaction is then given by
\begin{eqnarray}
\sigma_{aa\to ss}=\frac{1}{64\pi I^2} \int_{t_1}^{t_2} |\mathcal{M}_{aa\to ss}|^2 (s,t) dt \label{eqsaass}
\end{eqnarray}
where 
\begin{equation}
t_{1,2}=m^2+m_s^2-\frac{s}{2}\pm\sqrt{s}\sqrt{\frac{s}{4}-m_s^2}\,,
\end{equation}
and
\begin{equation}
I=\sqrt{\frac{2m^4 + s^2 -4sm^2}{2}}\,.
\end{equation}
The differential cross section for the production of a sterile neutrino with energy $E_s$ in the process $\nu+\nu\to \nu_s+\nu_s$ is
\begin{eqnarray}
\frac{d\sigma_{aa\to ss}}{dE_s}&=&\frac{|\mathcal{M}_{aa\to ss}|^2 [2pm, m^2+m_s^2 -2m(E-E_s)]}{32\pi E I} \times \nonumber \\ 
&\times&\theta\left(E-\frac{2mE_s^2}{2mE_s-m_s^2}\right) \theta\left(E_s-\frac{m_s^2}{2m}\right)\,.\label{eq12}
\end{eqnarray}
Here $E$ is the energy of the incident cosmogenic active neutrino.\\

Let us discuss now the second process $\nu+\nu_s\to \nu+\nu_s$. Again, the initial momentum of the background neutrino is $l$, the momentum of the incident sterile neutrino is $p$, the momentum of the final sterile and active neutrinos are respectively $k_1$ and $k_2$ as shown in Fig. (\ref{feyn2}). Notice that in this case, since the two final particles are distinguishable, the choice of how to define the Mandelstam parameters is not equivalent: we choose the convention that $t=(p-k_2)^2=(l-k_1)^2$. With this choice, the squared amplitude $|\mathcal{M}_{as\to as}|^2$  is identical to Eq.\,\eqref{amplit} with the $s$ and the $u$ parameters exchanged in the corresponding equation. The total cross section for the process is
\begin{eqnarray}
\sigma_{as\to as}=\frac{1}{64\pi J^2} \int_{t_1}^{t_2} |\mathcal{M}_{aa\to ss}|^2 (m_s^2+2mE,t) dt\nonumber\\\label{eqsasas}
\end{eqnarray}
with 
\begin{equation}
t_{1,2}=m^2+m_s^2 -\frac{(m_s^2+2mE)^2-m_s^4}{2(m_s^2+2mE)}\pm \frac{2m^2E^2}{2mE+m_s^2}
\end{equation}
 and again the energy of the incident sterile neutrino is $E$ and  $J$ is defined as
\begin{equation}
J=\sqrt{\frac{m^4 + m_s^4 +s^2-2sm^2-2sm_s^2}{2}}\,.
\end{equation}

The differential cross section for the production of an active neutrino of energy $E_2$ is then
\begin{eqnarray}\label{eq14a}
&&\frac{d\sigma_{as\to as}}{dE_2}=\frac{1}{32\pi E J} \theta\left(\frac{2mE^2}{2mE+m_s^2}-E_2\right) \times \\ 
&& \times |\mathcal{M}|^2 [m^2 +m_s^2 +2mE, m^2+m_s^2-2m(E-E_2)]\,.\nonumber
\end{eqnarray}
The differential cross section for the production of a sterile neutrino of energy $E_1$ is:
\begin{eqnarray}
&&\frac{d\sigma_{as\to as}}{dE_1}=\frac{1}{32\pi E J}\theta\left((E-E_1)(2mEE_1 - m_s^2 (E-E_1)\right)  \times \nonumber\\
&& \times |\mathcal{M}|^2 [m^2 + m_s^2 +2mE, m^2+m_s^2-2mE_1]\,.
\label{eq15}
\end{eqnarray}
Concerning the scalar mediator,  its decay rate is given by
\begin{equation}
\Gamma=\frac{\lambda^2 \xi (m m_s+\sqrt{\xi^2+m^2}\sqrt{\xi^2+m_s^2} + \xi^2)}{2\pi M_\varphi (\sqrt{\xi^2+m^2}+\sqrt{k^2+m_s^2})} \theta(M_\varphi-m-m_s)
\end{equation}
where
\begin{equation}
\xi=\frac{\sqrt{m^4-2m^2 M_\varphi^2 + M_\varphi^4 -2m^2 m_s^2 -2M_\varphi^2 m_s^2 +m_s^4}}{2M_\varphi}
\end{equation}
A point to emphasize is the fact that, for $m_s\geq M_\varphi$, the decay rate of the scalar mediator vanishes, since there is no decay channel kinematically allowed. This means that the resonances in the cross sections become unregulated. While this is not a problem for the $s$-resonance, which can never be reached in the physical space of parameters of the collision, the $t$- and $u$-resonance exhibit instead a singular behavior. This behavior needs to be regulated taking into account the finite transverse amplitude of the scattering beams, in a way analogous to \cite{Melnikov:1996iu}. In order to avoid this difficulty, we have restricted to the case $M_\varphi>m_s$.

\section{Propagation: transport equation}

The effect of the secret interaction on the neutrino flux produced through the $p\gamma$ interactions is described by a Boltzmann equation. The relevant physical processes are the collisions of a cosmogenic neutrino with a neutrino from the CNB, and two different effects are in principle possible: on the one hand, the collision of an astrophysical neutrino produces a depletion in the flux, described by an absorption term; on the other hand, after the collision two new daughter sterile neutrinos are produced. If the incident neutrinos are highly relativistic, as we are assuming, the collisions will be strongly forward, with collinear emission of the daughter neutrinos, and we can assume that in principle they will replenish the original flux. Nonetheless, we will see that, even though they are produced with the correct angle, their energies will be too low to be relevant to our work. The interplay between these two processes is described by the transport equation which we show below.

A subtle point which has to be taken into account is the effect of oscillations. In principle, we should write a differential evolution equation for each of the components of the density matrix in flavor space. However, at the extremely high energies of interest to us, the De Broglie wavelength of the neutrinos, which characterizes the distances over which neutrino oscillates, is much smaller than the characteristic distance of propagation, which is the mean free path for the interaction. With this consideration, we are assured that between two successive collision the oscillations have averaged out. This means that, even though neutrinos are produced as eigenstates of flavor, during their propagation their density matrix averages to a form which is diagonal in the space of the mass eigenstates. For this reason, we have studied the propagation equation for the fluxes of neutrinos in the mass eigenstates. 

Let $\Phi_a (z,E)$ be the flux of active neutrinos per unit energy interval per unit solid angle  at a redshift $z$, $\Phi_s$ the analogous flux of sterile neutrino: we will collectively denote them by $\Phi_l$ where $\Phi_l \equiv d\phi_l/ dE\,d\Omega$. The transport set of equations is:
\begin{eqnarray} \label{transport}
&&H(z)\left[\frac{\partial \Phi_l}{\partial z} +\frac{\partial \Phi_l}{\partial E} \frac{E}{1+z}\right] = n(z) \sigma_l(E) \Phi_l(E) + \nonumber \\ 
&& -n(z) \int dE' \sum_{m=a,s} \frac{d\sigma_{ml}}{dE} (E'\to E) \Phi_m (E') + \nonumber \\ 
&&-\rho(z) f(E) \delta_{la}\,\,\,\,\,\,\,\,\,\,  {\rm with} \, l=a,s
\end{eqnarray}
Here $\sigma_l$  is the total cross section for collision of a neutrino of type $l=a,s$  with a neutrino from the CNB,  where $\sigma_a\equiv \sigma_{aa\to ss}$  of Eq.\,(\ref{eqsaass}) and $\sigma_s\equiv \sigma_{as\to as}$  of  Eq.\,(\ref{eqsasas}) (we are assuming a single active flavor for the latter, since the CNB is composed only of active neutrinos).\\ Moreover in Eq.\,(\ref{transport}) the quantity $\frac{d\sigma_{ml}}{dE} (E'\to E)$ denotes the partial cross section of  Eq.s\,(\ref{eq14a}) and (\ref{eq15})
for the production of an $l$-th neutrino with energy $E$ after the collision of an $m$-th neutrino of energy $E'$ with the CNB one. The quantity $n$ denotes the number density of CNB neutrinos, which we have taken to be $n(z)=n_0 (1+z)^3$ with $n_0=116 \rm{cm}^{-3}$. The function $f(E)$ is the number of neutrinos emitted per unit energy interval per unit time per unit solid angle, which has been described above.  $\rho(z)$ is the density of sources which has been taken to evolve with the Star Formation Rate.

Equation \eqref{transport} is a system of two partial differential coupled equations, which should in principle be solved numerically. However, some physical considerations allow us to obtain the most interesting results with a simplified approach. If there is no mixing from oscillations between the active and the sterile neutrinos, then we have no interest in the sterile flux at Earth, which could not be detected in any case. In the equation for the active flux, the term describing the replenishment of the flux by the process $\nu_s+\nu\to \nu_s+\nu$ is weighted by the differential cross section for production of an active neutrino. We will now describe the order of magnitude of this term. If $l$ is the order of magnitude of the distance traveled by the neutrino, which can be taken to be $10^{26}$ m, the correction to the active flux is of order:
\begin{equation}
\Delta\Phi_a (E) \sim n l \int dE' \Phi_s (E') \frac{d\sigma_{sa}}{dE} (E'\to E)
\end{equation}
The sterile flux is generated by the active flux itself, and can be estimated in the same way, obtaining:
\begin{eqnarray} \label{new}
\Delta\Phi_a(E) \sim n^2 l^2  \int dE' \int dE'' \Phi_a (E'') \times \nonumber \\ \times \frac{d\sigma_{as}}{dE'} (E''\to E') \frac{d\sigma_{sa}}{dE} (E'\to E)
\end{eqnarray}
The mean value of the energy $E''$, due to the kinematic threshold for the interaction of active neutrinos set at $2 m_s^2/m$, which is at least of order $10^9$ GeV, turns out to be very large, at least of order $10^{10}$ GeV. In other words an active neutrino can be produced through regeneration by a sterile neutrino, which has to be produced itself by an active neutrino. The latter has to have an energy at least as high as $10^{10}$ GeV. Due to the rapid decrease with energy of the input flux, this correction turns out to be much smaller than the original flux. Further, we have numerically solved the equation for some benchmark cases, finding in fact that for decreasing fluxes, as in our case, the correction for regeneration is irrelevant, while it could be relevant in case of increasing fluxes.

Therefore, we can neglect the regeneration term in the equation for the active neutrinos, which becomes:
\begin{eqnarray}
&&H(z)\left[\frac{\partial \Phi_a}{\partial z} +\frac{\partial \Phi_a}{\partial E} \frac{E}{1+z}\right] = n(z) \sigma_a(E) \Phi_a(E) + \nonumber \\ 
&&-\rho(z) f(E)
\end{eqnarray}
This equation contains only an absorption term, and admits now an analytical solution for the flux at Earth:
\begin{eqnarray} \label{abs}
\Phi_a (E)=\int_0^{+\infty} \frac{dz}{H(z)} \rho(z) f\left[ E(1+z)\right] \times \nonumber \\ \times \exp\left[-\int_0^z \frac{dz'}{H(z')} n(z') \sigma_a\left[E(1+z')\right]\right]
\end{eqnarray}
The validity of this approximation has been verified by explicitly finding the numerical solution to the full system \eqref{transport} for some benchmark values of the sterile mass and the mediator mass between $250$ MeV and $1$ GeV, and comparing it to \eqref{abs}. We found a perfect agreement between the two.

\section{Constraints}

The model we are assuming is in principle subject to a number of constraints from cosmology, astrophysics and laboratory experiments. We did not perform a detailed study of the unconstrained region spanned by the parameters  $\lambda,\, m_s,\, M$, since this would be beyond the scope of this work. We focused on a particular region which turns out not to be affected by these constraints.\\

\noindent{\it Laboratory bounds}\\
 Mesons decays, in particular Kaon decays, can be quite restrictive for models with secret interactions, since the latter can introduce new decay channels. For secret interactions among active neutrinos severe constraints come from meson decays in the $(\lambda,M)$ plane, as shown for instance in \cite{Berryman:2018ogk} and \cite{deGouvea:2019qaz}, where it is roughly found that the region up to $M \gtrsim 250\,\mbox{MeV}$ is insensitive to the exclusions. For our model one should in principle make a similar analysis, since through the secret interaction the kaon can decay into $K\to \mu\nu_s\varphi$ or $K\to\mu\nu_s\nu_s\nu$. If, however, we restrict to the range of masses $m_s\geq 250$ MeV and $M \geq 250$ MeV, these reactions are kinematically forbidden, justifying the absence of constraints from these decays.\\ \\

\noindent{\it Big Bang Nucleosyntesis bounds}\\
It is interesting to note that with the choice $M, m_s\gtrsim 250$ MeV, the whole sector of the active and sterile neutrino and the scalar mediator remains in equilibrium at temperatures even lower than the BBN. For example, if $\lambda=1$, the interactions are relevant down to temperatures of $100$ eV. Of course, after around $1$ MeV, the active neutrino decouples from the Standard Model plasma, so below this temperature we cannot speak of thermal equilibrium, but the interactions between active and sterile neutrinos and scalar mediator remain effective. Because of this, the sterile neutrino and the scalar mediator are Boltzmann suppressed at temperatures below essentially $T\sim 100$ MeV, since at these temperatures the reactions of production of sterile neutrinos are kinematically suppressed. Therefore at the BBN temperatures, around $1$ MeV, the newly introduced particles have disappeared and they do not count as radiative degrees of freedom, without in fact influencing the nucleosynthesis. We also point out that, as shown in \cite{Grohs:2020xxd}, there is another possible imprint that our interaction might leave on the BBN, due to the presence of an active-active interaction which may distort the Fermi distribution of active neutrinos. However, the active-active interaction can only happen either through mixing or through next-to-leading order interactions, since our model Lagrangian only contains an active-sterile vertex. Assuming next-to-leading order interactions are sufficiently suppressed, the rate for active-active interactions through mixing is $n\sigma v\sim T^3 \frac{\lambda^4 \theta^4 T^2}{M_\phi^4}$, where $\theta$ is the active-sterile mixing angle. This interaction is relevant only until the point at which $H\sim T^2/M_{\rm{Pl}}$, where $H$ is the Hubble parameter and $M_{\rm{Pl}}$ is the Planck mass. Equating these we find that for $\lambda=1$, $M_{\phi}=250$ MeV and $\theta\leq 10^{-4}$ the active-active interaction is irrelevant at the moment of the BBN.\\

\noindent{\it Cosmic Microwave Background bounds}\\
At the time of formation of the Cosmic Microwave Background, the sterile and scalar particles have long disappeared. The active neutrinos can interact through the four point reactions $\nu\nu \to \nu\nu $. As mentioned above, this interaction should have already run out of equilibrium before the BBN, to avoid changes in the distribution functions of the neutrinos, so it is even less relevant at the time of the formation of the CMB.\\ 

\noindent{\it Astrophysical bounds}\\
Another constraint might in principle come from the analysis of neutrino fluxes from supernovae. In fact, since neutrinos in the core of supernovae have energies of order of tenth or hundredth of MeV, they are sufficiently energetic to produce sterile neutrinos which could escape the supernova, giving rise to an energy loss with observable consequences. However, due to the interactions we are introducing, even though sterile neutrinos can in principle be produced, they are not able to escape the supernova, trapped by the secret interaction with the active neutrinos inside the core. In order to verify this statement, we have computed the order of magnitude of the mean free path of a sterile neutrino inside the core. We have obtained, for example, for a benchmark mass of $250$ MeV for the sterile neutrino, $300$ MeV for the scalar mediator and $1$ for the coupling, a mean free path of $10^{-11}$ m, clearly much smaller than the characteristic distances of a supernova.

In view of the above considerations we restrict in the following our analysis to the case of massive sterile neutrinos, with $m_s$ of few hundreds MeV and a mass of the pseudoscalar mediator $M_\varphi  \geq m_s$.

\section{Results and detection chances}

Among the most relevant experiments that have performed a detection campaign for cosmogenic neutrinos, and the future apparatus that will undertake such measurements, we have taken into account, as representatives of different classes, the sensitivities of PAO, GRAND and ARIANNA for comparison.


The main goal of PAO, located in Argentina and taking data since 2004, was the measurement of extensive air showers produced by secondary particles in the interaction of UHECRs with the atmosphere. The Pierre Auger Observatory consists of two different experimental set up performing two independent measurements of the shower: a Fluorescence Detector (FD) and a Surface Detector (SD). The FD detects the Nitrogen fluorescence emission produced during the development of the shower in the atmosphere by means of $24$ large telescopes placed at four observation sites located atop small elevations on the perimeter of the SD array. It is sensitive to UHECRs with energy above ($10^{18}$ eV) . On the other side, the SD apparatus consists of an array of water-Cherenkov detectors that are located on a large area of about 3000 $\rm{km}^2$ arranged in a hexagonal pattern.

Although the primary goal of PAO was to study UHECRs, it has been shown in \cite{Berezinsky:1975zz} that it can also study cosmogenic neutrinos.  Indeed neutrinos arriving at large zenith angle (horizontal with respect to the detector) produce at the sea level extensive air showers with small radius of curvature \cite{Berezinsky:1975zz}, in contrast with other showers from large zenith angles. The electromagnetic component of ordinary air showers at large zenith angles   from hadronic cosmic rays is attenuated by the atmosphere before arriving at the sea level. Deeply penetrating particles like neutrinos come instead unattenuated \cite{Capelle:1998zz}. Recent upgrades of the PAO sensitivity can be found in \cite{Aab:2019auo} and an integrated sensitivity to cosmogenic neutrino fluxes is found of  about $4\cdot 10^{-9}\, \mbox{GeV} \,cm^{-2}\, s^{-1}\, sr^{-1}$.  Such a sensitivity will be improved by the Giant Radio Array for Neutrino Detection (GRAND)  \cite{Alvarez-Muniz:2018bhp}, that will be located in various favorable mountainous places in the world and is planned to take data in 2025 and should reach, after $10$ years of data, an integrated sensitivity to cosmogenic neutrino fluxes of $1\cdot 10^{-10}\, \mbox{GeV} \,cm^{-2}\, s^{-1}\, sr^{-1}$, thus improving of a factor of ten the current PAO sensitivity. GRAND will detect the radio emission coming from large particle showers, namely extensive air showers, like PAO. 
In the first detection stage,  GRAND10k  will use an array of 10.000 radio antennas deployed over an area of 10.000 km2 GRAND10k. A second stage for grand is planned, GRAND200k with 200.000 receivers, which could take data starting from 2030.

A further class of experiments are the Askaryan radio experiments, like ARIANNA \cite{Nelles:2018gqq}.
The ARIANNA experiment, located in the South Pole, aims to detect the radio signals of cosmogenic neutrinos. The ARIANNA concept is based on installing high-gain log periodic dipole antennas close to the surface monitoring the underlying ice for the radio signals following a neutrino interaction and based on the Askaryan effect.

\begin{figure}[h]
\begin{center}
\includegraphics[width=0.45\textwidth]{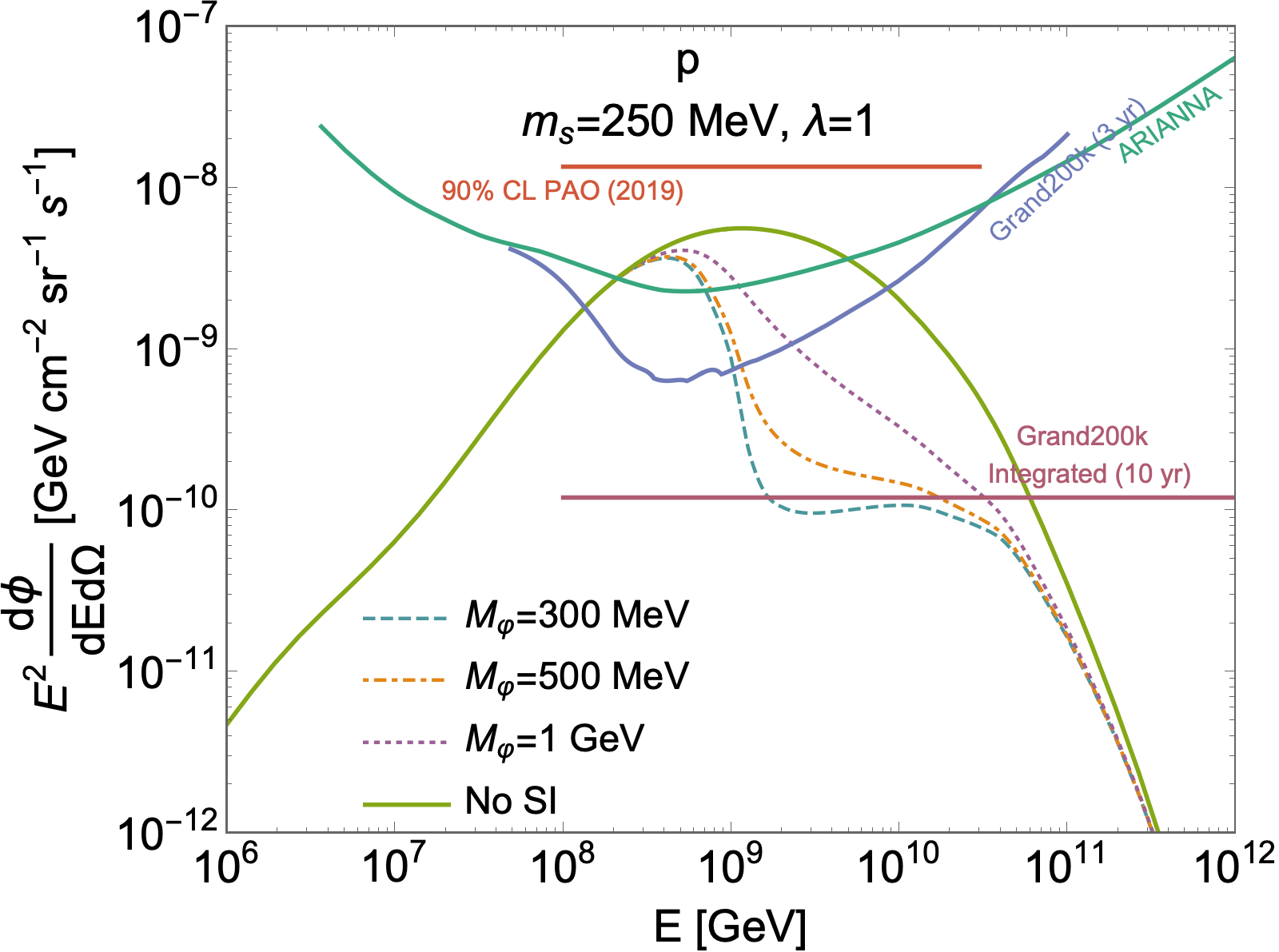}
\includegraphics[width=0.45\textwidth]{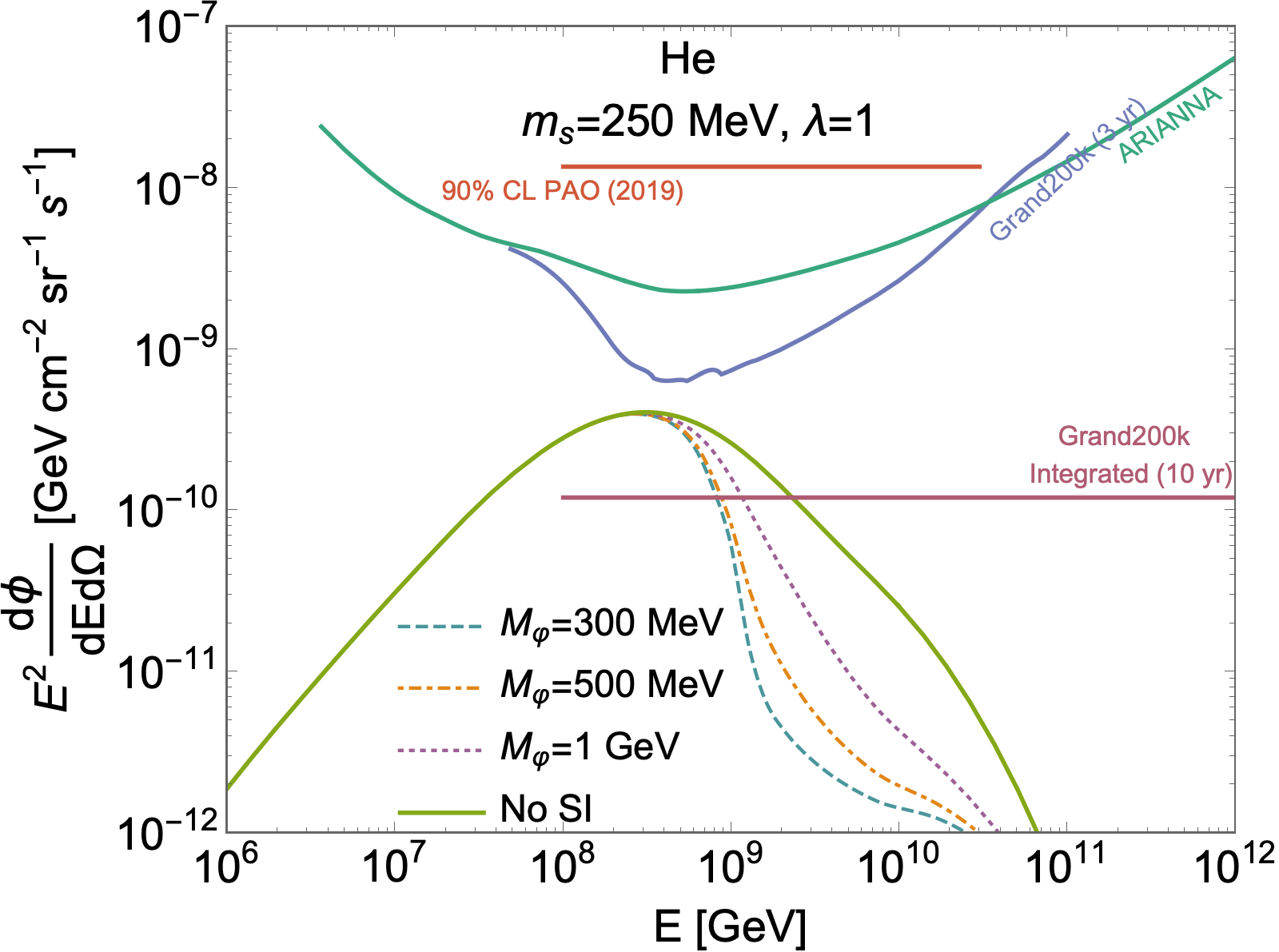}
\caption{Effects on the cosmogenic spectrum expected in the case of: {\it top panel)} proton cosmic rays,  {\it bottom panel)} helium cosmic rays. The continuous green curve is the cosmogenic spectrum expected in the absence of secret interactions, the dashed, dotdashed and dotted lines are the spectrum for secret interactions with $m_s=250$ MeV and  $M_\varphi = 300, 500, 1000$ MeV  respectively. The sensitivity of the GRAND experiment, the 90\% C.L. of PAO, the integrated sensitivity of GRAND after 10 years of data and the sensitivity of the ARIANNA experiment are also shown.}\label{fig1}
\end{center}
\end{figure}

In Figure \ref{fig1} we provide the expected cosmogenic spectra both in the absence of secret interactions and in their presence, for some benchmark values of the scalar mediator masses. These spectra have been obtained under the assumption of a purely protonic UHECRs flux (top panel) and purely helium cosmic rays (bottom panel). The coupling has been fixed to $\lambda=1$, while the sterile mass is taken to be $250$ MeV. The dashed, dotdashed and dotted curve respectively correspond to a mass of the scalar mediator of $300$ MeV, $500$ MeV and $1$ GeV. In agreement with the expectations, larger masses for the mediator correspond to weaker absorption.

The sensitivities of the GRAND, the PAO and the ARIANNA \cite{Barwick:2016cyg} experiments are shown as well. It appears that already the GRAND and the ARIANNA experiment after $3$ years of data taking would be able to distinguish the presence of the secret interaction, even for masses as large as $500$ MeV, at least in the case of purely protonic UHECRs. In fact, while in the absence of the secret interaction the cosmogenic flux should be detected up to energies of $\sim 10^{10}$ GeV, being above the sensitivity of the experiment, the interaction causes its absorption, making the flux undetectable already at energies of $\sim 10^9$ GeV.

This conclusion depends very strongly, however, on our choice for the coupling. In fact, since the cross section, appearing in the exponent of the absorption coefficient, grows as the fourth power of the coupling, already a choice of $\lambda\sim 0.1$ renders the absorption effect completely negligible. A milder dependence is the one expected on the sterile neutrino masses. Raising this mass causes an increase in the threshold energy for the production. In fact, the latter is $E_{\rm{th}}\simeq m_s^2/m$, which is the energy at which we expect the absorption effect to begin. 

All of our conclusions are of course based on the assumption of a purely protonic UHECRs flux. Recent data \cite{Castellina:2019huz} suggest a mixed chemical composition, tending towards Helium and Nitrogen as the dominant components. It is therefore of interest to determine how our results change for a different chemical composition. In the case of Helium dominated cosmic rays, the spectrum without interaction is already lower than the GRAND sensitivity at $3$ years, while it is slightly higher than the integrated sensitivity after $10$ years of data taking.

For Nitrogen dominated cosmic rays, the spectrum, with or without secret interaction, is not amenable to detection at GRAND, being even below the integrated sensitivity at $10$ years. Since our partial informations about the chemical composition of UHECRs suggest a mixed composition, it is reasonable to expect a cosmogenic spectrum somewhere in between the extreme cases analyzed.

Nonetheless, even in the pessimistic case of extremely low cosmogenic neutrino fluxes due to heavy ion dominance of UHECRs, the same effect might be relevant for neutrinos from astrophysical sources. In fact, neutrinos from blazars can reach energies as high as $10^{10}$ GeV \cite{Murase:2014foa}, whereas the position of the resonant absorption is around $\sim 10^8$ GeV.

\section{Conclusions}
In this paper we have investigated the experimentally observable effects coming from a simple form of active-sterile secret interactions in neutrino sector, mediated by a pseudoscalar,  at air shower experiments like GRAND and PAO. Due to the presence of such interactions, the flux of cosmogenic neutrinos results to be strongly distorted.
In this scenario, in order to evade the constraints coming from cosmology, astrophysics and particle physics, we need to take quite a large mass for the sterile neutrinos of the order of few hundreds MeV. Observable effects are then seen in the cosmogenic neutrino fluxes of ultrahigh energy, produced through the GZK interaction. In fact, these extremely energetic neutrinos can, during their path to Earth, collide with neutrinos from the Cosmic Neutrino Background, producing a flux of sterile neutrinos and depleting the flux of active ones. We showed that this results in an absorption of the cosmogenic neutrino fluxes and then in a peculiar distortion of the expected flux. We also showed that, in the case of GZK neutrinos coming from cosmic rays dominated by protons, the effects of these absorptions are sufficiently important to be revealed by a forthcoming experiment like GRAND or ARIANNA. In the case of different chemical compositions of the cosmic rays, the neutrino flux is generally lower even without absorption and the possibility of cosmogenic neutrino detection with the GRAND experiment is only marginal. In this case, the cosmogenic neutrino detection would demand for the future a much larger generation of cosmic rays apparatus. Even in such a more pessimistic case, the effect we have investigated can still be relevant for neutrino fluxes from astrophysical sources.

\vskip10.mm

{\it Acknowledgments:} we thanks Elisa Bernardini, Alessandro Mirizzi and Kohta Murase for useful discussion.  This work was partially supported by the research grant number 2017W4HA7S ``NAT-NET: Neutrino and Astroparticle Theory Network'' under the program PRIN 2017 funded by the Italian Ministero dell'Universit\`a e della Ricerca (MUR). The authors acknowledge partial support by the research project TAsP (Theoretical Astroparticle Physics) funded by the Instituto Nazionale di Fisica Nucleare (INFN).

\bibliographystyle{unsrt}

\begin{thebibliography}{10}

\bibitem{Aartsen:2013jdh}
M.~G. Aartsen et~al.
\newblock {Evidence for High-Energy Extraterrestrial Neutrinos at the IceCube
  Detector}.
\newblock {\em Science}, 342:1242856, 2013.

\bibitem{Beresinsky:1969qj}
V.~S. Berezinsky and G.~T. Zatsepin.
\newblock {Cosmic rays at ultrahigh-energies (neutrino?)}.
\newblock {\em Phys. Lett.}, 28B:423--424, 1969.

\bibitem{Greisen:1966jv}
Kenneth Greisen.
\newblock {End to the cosmic ray spectrum?}
\newblock {\em Phys. Rev. Lett.}, 16:748--750, 1966.

\bibitem{Zatsepin:1966jv}
G.~T. Zatsepin and V.~A. Kuzmin.
\newblock {Upper limit of the spectrum of cosmic rays}.
\newblock {\em JETP Lett.}, 4:78--80, 1966.
\newblock [Pisma Zh. Eksp. Teor. Fiz.4,114(1966)].

\bibitem{Berezinsky:1998ft}
Veniamin Berezinsky, Pasquale Blasi, and Alexander Vilenkin.
\newblock {Ultrahigh-energy gamma-rays as signature of topological defects}.
\newblock {\em Phys. Rev.}, D58:103515, 1998.

\bibitem{Nagano:2000ve}
M.~Nagano and Alan~A. Watson.
\newblock {Observations and implications of the ultrahigh-energy cosmic rays}.
\newblock {\em Rev. Mod. Phys.}, 72:689--732, 2000.

\bibitem{Engel:2001hd}
Ralph Engel, David Seckel, and Todor Stanev.
\newblock {Neutrinos from propagation of ultrahigh-energy protons}.
\newblock {\em Phys. Rev.}, D64:093010, 2001.

\bibitem{Kusenko:2001gj}
Alexander Kusenko and Thomas~J. Weiler.
\newblock {Neutrino cross-sections at high-energies and the future observations
  of ultrahigh-energy cosmic rays}.
\newblock {\em Phys. Rev. Lett.}, 88:161101, 2002.

\bibitem{Anchordoqui:2002hs}
Luis Anchordoqui, Thomas~Cantzon Paul, Stephen Reucroft, and John Swain.
\newblock {Ultrahigh-energy cosmic rays: The State of the art before the Auger
  Observatory}.
\newblock {\em Int. J. Mod. Phys.}, A18:2229--2366, 2003.

\bibitem{Fodor:2002hy}
Z.~Fodor, S.~D. Katz, and A.~Ringwald.
\newblock {Relic neutrino masses and the highest energy cosmic rays}.
\newblock {\em JHEP}, 06:046, 2002.

\bibitem{Kalashev:2002kx}
Oleg~E. Kalashev, Vadim~A. Kuzmin, Dmitry~V. Semikoz, and Gunter Sigl.
\newblock {Ultrahigh-energy neutrino fluxes and their constraints}.
\newblock {\em Phys. Rev.}, D66:063004, 2002.

\bibitem{Semikoz:2003wv}
Dmitry~V. Semikoz and Gunter Sigl.
\newblock {Ultrahigh-energy neutrino fluxes: New constraints and implications}.
\newblock {\em JCAP}, 0404:003, 2004.

\bibitem{Fodor:2003ph}
Z.~Fodor, S.~D. Katz, A.~Ringwald, and H.~Tu.
\newblock {Bounds on the cosmogenic neutrino flux}.
\newblock {\em JCAP}, 0311:015, 2003.

\bibitem{Ave:2004uj}
Maximo Ave, N.~Busca, Angela~V. Olinto, Alan~A. Watson, and T.~Yamamoto.
\newblock {Cosmogenic neutrinos from ultra-high energy nuclei}.
\newblock {\em Astropart. Phys.}, 23:19--29, 2005.

\bibitem{Seckel:2005cm}
David Seckel and Todor Stanev.
\newblock {Neutrinos: The Key to UHE cosmic rays}.
\newblock {\em Phys. Rev. Lett.}, 95:141101, 2005.

\bibitem{DeMarco:2005kt}
Daniel De~Marco, Todor Stanev, and F.~W. Stecker.
\newblock {Cosmogenic neutrinos from cosmic ray interactions with extragalactic
  infrared photons}.
\newblock {\em Phys. Rev.}, D73:043003, 2006.

\bibitem{Allard:2006mv}
Denis Allard, M.~Ave, N.~Busca, M.~A. Malkan, A.~V. Olinto, E.~Parizot, F.~W.
  Stecker, and T.~Yamamoto.
\newblock {Cosmogenic Neutrinos from the propagation of Ultrahigh Energy
  Nuclei}.
\newblock {\em JCAP}, 0609:005, 2006.

\bibitem{Becker:2007sv}
Julia~K. Becker.
\newblock {High-energy neutrinos in the context of multimessenger physics}.
\newblock {\em Phys. Rept.}, 458:173--246, 2008.

\bibitem{Anchordoqui:2007fi}
Luis~A. Anchordoqui, Haim Goldberg, Dan Hooper, Subir Sarkar, and Andrew~M.
  Taylor.
\newblock {Predictions for the Cosmogenic Neutrino Flux in Light of New Data
  from the Pierre Auger Observatory}.
\newblock {\em Phys. Rev.}, D76:123008, 2007.

\bibitem{Berezinsky:2010xa}
V.~Berezinsky, A.~Gazizov, M.~Kachelriess, and S.~Ostapchenko.
\newblock {Restricting UHECRs and cosmogenic neutrinos with Fermi-LAT}.
\newblock {\em Phys. Lett.}, B695:13--18, 2011.

\bibitem{Ahlers:2010fw}
M.~Ahlers, L.~A. Anchordoqui, M.~C. Gonzalez-Garcia, F.~Halzen, and S.~Sarkar.
\newblock {GZK Neutrinos after the Fermi-LAT Diffuse Photon Flux Measurement}.
\newblock {\em Astropart. Phys.}, 34:106--115, 2010.

\bibitem{Katz:2011ke}
U.~F. Katz and Ch. Spiering.
\newblock {High-Energy Neutrino Astrophysics: Status and Perspectives}.
\newblock {\em Prog. Part. Nucl. Phys.}, 67:651--704, 2012.

\bibitem{Gelmini:2011kg}
Graciela~B. Gelmini, Oleg Kalashev, and Dmitri~V. Semikoz.
\newblock {Gamma-Ray Constraints on Maximum Cosmogenic Neutrino Fluxes and
  UHECR Source Evolution Models}.
\newblock {\em JCAP}, 1201:044, 2012.

\bibitem{Ng:2014pca}
Kenny C.~Y. Ng and John~F. Beacom.
\newblock {Cosmic neutrino cascades from secret neutrino interactions}.
\newblock {\em Phys. Rev.}, D90(6):065035, 2014.
\newblock [Erratum: Phys. Rev.D90,no.8,089904(2014)].

\bibitem{Murase:2014tsa}
Kohta Murase.
\newblock {On the Origin of High-Energy Cosmic Neutrinos}.
\newblock {\em AIP Conf. Proc.}, 1666(1):040006, 2015.

\bibitem{Heinze:2015hhp}
Jonas Heinze, Denise Boncioli, Mauricio Bustamante, and Walter Winter.
\newblock {Cosmogenic Neutrinos Challenge the Cosmic Ray Proton Dip Model}.
\newblock {\em Astrophys. J.}, 825(2):122, 2016.

\bibitem{Aloisio:2015ega}
R.~Aloisio, D.~Boncioli, A~di~Matteo, A.~F. Grillo, S.~Petrera, and
  F.~Salamida.
\newblock {Cosmogenic neutrinos and ultra-high energy cosmic ray models}.
\newblock {\em JCAP}, 1510(10):006, 2015.

\bibitem{Halzen:2016gng}
Francis Halzen.
\newblock {High-energy neutrino astrophysics}.
\newblock {\em Nature Phys.}, 13(3):232--238, 2016.

\bibitem{Cherry:2018rxj}
John~F. Cherry and Ian~M. Shoemaker.
\newblock {Sterile neutrino origin for the upward directed cosmic ray showers
  detected by ANITA}.
\newblock {\em Phys. Rev.}, D99(6):063016, 2019.

\bibitem{Vitagliano:2019yzm}
Edoardo Vitagliano, Irene Tamborra, and Georg Raffelt.
\newblock {Grand Unified Neutrino Spectrum at Earth}.
\newblock 2019.

\bibitem{Abbasi:2007sv}
R.~U. Abbasi et~al.
\newblock {First observation of the Greisen-Zatsepin-Kuzmin suppression}.
\newblock {\em Phys. Rev. Lett.}, 100:101101, 2008.

\bibitem{Abraham:2004dt}
J.~Abraham et~al.
\newblock {Properties and performance of the prototype instrument for the
  Pierre Auger Observatory}.
\newblock {\em Nucl. Instrum. Meth.}, A523:50--95, 2004.

\bibitem{Abraham:2008ru}
J.~Abraham et~al.
\newblock {Observation of the suppression of the flux of cosmic rays above
  $4\times 10^{19}$eV}.
\newblock {\em Phys. Rev. Lett.}, 101:061101, 2008.

\bibitem{Alvarez-Muniz:2018bhp}
Jaime Alvarez~Muniz et~al.
\newblock {The Giant Radio Array for Neutrino Detection (GRAND): Science and
  Design}.
\newblock {\em Sci. China Phys. Mech. Astron.}, 63(1):219501, 2020.

\bibitem{Anker:2019rzo} 
  A.~Anker {\it et al.},
  \newblock{A search for cosmogenic neutrinos with the ARIANNA test bed using 4.5 years of data}
  \newblock{arXiv:1909.00840 [astro-ph.IM].}
  
  \bibitem{Allison:2015eky} 
  P.~Allison {\it et al.} [ARA Collaboration],
  \newblock{Performance of two Askaryan Radio Array stations and first results in the search for ultrahigh energy neutrinos}
  \newblock{Phys.\ Rev.\ D} {\bf 93}, no. 8, 082003 (2016)

\bibitem{Adams:2013vea}
J.~H. Adams et~al.
\newblock {An evaluation of the exposure in nadir observation of the JEM-EUSO
  mission}.
\newblock {\em Astropart. Phys.}, 44:76--90, 2013.

\bibitem{Ahlers:2012rz}
Markus Ahlers and Francis Halzen.
\newblock {Minimal Cosmogenic Neutrinos}.
\newblock {\em Phys. Rev.}, D86:083010, 2012.

\bibitem{Castellina:2019huz}
Antonella Castellina.
\newblock {Highlights from the Pierre Auger Observatory (ICRC2019)}.
\newblock {\em PoS}, ICRC2019:004, 2020.

\bibitem{Davidson:2003ha}
S.~Davidson, C.~Pena-Garay, N.~Rius, and A.~Santamaria.
\newblock {Present and future bounds on nonstandard neutrino interactions}.
\newblock {\em JHEP}, 03:011, 2003.

\bibitem{Antusch:2008tz}
Stefan Antusch, Jochen~P. Baumann, and Enrique Fernandez-Martinez.
\newblock {Non-Standard Neutrino Interactions with Matter from Physics Beyond
  the Standard Model}.
\newblock {\em Nucl. Phys.}, B810:369--388, 2009.

\bibitem{Miranda:2004nb}
O.~G. Miranda, M.~A. Tortola, and J.~W.~F. Valle.
\newblock {Are solar neutrino oscillations robust?}
\newblock {\em JHEP}, 10:008, 2006.

\bibitem{Fornengo:2001pm}
N.~Fornengo, M.~Maltoni, R.~Tomas, and J.~W.~F. Valle.
\newblock {Probing neutrino nonstandard interactions with atmospheric neutrino
  data}.
\newblock {\em Phys. Rev.}, D65:013010, 2002.

\bibitem{Huber:2001zw}
Patrick Huber and J.~W.~F. Valle.
\newblock {Nonstandard interactions: Atmospheric versus neutrino factory
  experiments}.
\newblock {\em Phys. Lett.}, B523:151--160, 2001.

\bibitem{Barranco:2005ps}
J.~Barranco, O.~G. Miranda, C.~A. Moura, and J.~W.~F. Valle.
\newblock {Constraining non-standard interactions in nu(e) e or anti-nu(e) e
  scattering}.
\newblock {\em Phys. Rev.}, D73:113001, 2006.

\bibitem{Farzan:2017xzy}
Y.~Farzan and M.~Tortola.
\newblock {Neutrino oscillations and Non-Standard Interactions}.
\newblock {\em Front.in Phys.}, 6:10, 2018.

\bibitem{Ribeiro:2007ud}
N.~C. Ribeiro, H.~Minakata, H.~Nunokawa, S.~Uchinami, and
  R.~Zukanovich-Funchal.
\newblock {Probing Non-Standard Neutrino Interactions with Neutrino Factories}.
\newblock {\em JHEP}, 12:002, 2007.

\bibitem{Coloma:2015kiu}
Pilar Coloma.
\newblock {Non-Standard Interactions in propagation at the Deep Underground
  Neutrino Experiment}.
\newblock {\em JHEP}, 03:016, 2016.

\bibitem{deGouvea:2015ndi}
Andre de~Gouvea and Kevin~J. Kelly.
\newblock {Non-standard Neutrino Interactions at DUNE}.
\newblock {\em Nucl. Phys.}, B908:318--335, 2016.

\bibitem{Forero:2011pc}
D.~V. Forero, S.~Morisi, M.~Tortola, and J.~W.~F. Valle.
\newblock {Lepton flavor violation and non-unitary lepton mixing in low-scale
  type-I seesaw}.
\newblock {\em JHEP}, 09:142, 2011.

\bibitem{Mangano:2006ar}
Gianpiero Mangano, Gennaro Miele, Sergio Pastor, Teguayco Pinto, Ofelia
  Pisanti, and Pasquale~D. Serpico.
\newblock {Effects of non-standard neutrino-electron interactions on relic
  neutrino decoupling}.
\newblock {\em Nucl. Phys.}, B756:100--116, 2006.

\bibitem{Ioka:2014kca}
Kunihto Ioka and Kohta Murase.
\newblock {IceCube PeV?EeV neutrinos and secret interactions of neutrinos}.
\newblock {\em PTEP}, 2014(6):061E01, 2014.

\bibitem{Bakhti:2018avv}
P.~Bakhti, Y.~Farzan, and M.~Rajaee.
\newblock {Secret interactions of neutrinos with light gauge boson at the DUNE
  near detector}.
\newblock {\em Phys. Rev.}, D99(5):055019, 2019.

\bibitem{Kolb:1987qy}
Edward~W. Kolb and Michael~S. Turner.
\newblock {Supernova SN 1987a and the Secret Interactions of Neutrinos}.
\newblock {\em Phys. Rev.}, D36:2895, 1987.

\bibitem{Archidiacono:2013dua}
Maria Archidiacono and Steen Hannestad.
\newblock {Updated constraints on non-standard neutrino interactions from
  Planck}.
\newblock {\em JCAP}, 1407:046, 2014.

\bibitem{Laha:2013xua}
Ranjan Laha, Basudeb Dasgupta, and John~F. Beacom.
\newblock {Constraints on New Neutrino Interactions via Light Abelian Vector
  Bosons}.
\newblock {\em Phys. Rev.}, D89(9):093025, 2014.

\bibitem{Forastieri:2019cuf}
Francesco Forastieri, Massimiliano Lattanzi, and Paolo Natoli.
\newblock {Cosmological constraints on neutrino self-interactions with a light
  mediator}.
\newblock {\em Phys. Rev.}, D100(10):103526, 2019.

\bibitem{Bustamante:2020mep}
Mauricio Bustamante, Charlotte~Amalie Rosenstroem, Shashank Shalgar, and Irene
  Tamborra.
\newblock {Bounds on secret neutrino interactions from high-energy
  astrophysical neutrinos}.
\newblock 2020.

\bibitem{Blum:2014ewa}
Kfir Blum, Anson Hook, and Kohta Murase.
\newblock {High energy neutrino telescopes as a probe of the neutrino mass
  mechanism}.
\newblock 2014.

\bibitem{Murase:2019xqi}
Kohta Murase and Ian~M. Shoemaker.
\newblock {Neutrino Echoes from Multimessenger Transient Sources}.
\newblock {\em Phys. Rev. Lett.}, 123(24):241102, 2019.

\bibitem{Babu:2019iml}
K.~S. Babu, Garv Chauhan, and P.~S. Bhupal~Dev.
\newblock {Neutrino Non-Standard Interactions via Light Scalars in Earth, Sun,
  Supernovae and Early Universe}.
\newblock 2019.

\bibitem{Hannestad:2013ana}
Steen Hannestad, Rasmus~Sloth Hansen, and Thomas Tram.
\newblock {How Self-Interactions can Reconcile Sterile Neutrinos with
  Cosmology}.
\newblock {\em Phys. Rev. Lett.}, 112(3):031802, 2014.

\bibitem{Dasgupta:2013zpn}
Basudeb Dasgupta and Joachim Kopp.
\newblock {Cosmologically Safe eV-Scale Sterile Neutrinos and Improved Dark
  Matter Structure}.
\newblock {\em Phys. Rev. Lett.}, 112(3):031803, 2014.

\bibitem{Archidiacono:2014nda}
Maria Archidiacono, Steen Hannestad, Rasmus~Sloth Hansen, and Thomas Tram.
\newblock {Cosmology with self-interacting sterile neutrinos and dark matter -
  A pseudoscalar model}.
\newblock {\em Phys. Rev.}, D91(6):065021, 2015.

\bibitem{Saviano:2014esa}
Ninetta Saviano, Ofelia Pisanti, Gianpiero Mangano, and Alessandro Mirizzi.
\newblock {Unveiling secret interactions among sterile neutrinos with big-bang
  nucleosynthesis}.
\newblock {\em Phys. Rev.}, D90(11):113009, 2014.

\bibitem{Mirizzi:2014ama}
Alessandro Mirizzi, Gianpiero Mangano, Ofelia Pisanti, and Ninetta Saviano.
\newblock {Collisional production of sterile neutrinos via secret interactions
  and cosmological implications}.
\newblock {\em Phys. Rev.}, D91(2):025019, 2015.

\bibitem{Cherry:2014xra}
John~F. Cherry, Alexander Friedland, and Ian~M. Shoemaker.
\newblock {Neutrino Portal Dark Matter: From Dwarf Galaxies to IceCube}.
\newblock 2014.

\bibitem{Chu:2015ipa}
Xiaoyong Chu, Basudeb Dasgupta, and Joachim Kopp.
\newblock {Sterile neutrinos with secret interactions?lasting friendship with
  cosmology}.
\newblock {\em JCAP}, 1510(10):011, 2015.

\bibitem{Archidiacono:2016kkh}
Maria Archidiacono, Stefano Gariazzo, Carlo Giunti, Steen Hannestad, Rasmus
  Hansen, Marco Laveder, and Thomas Tram.
\newblock {Pseudoscalar?sterile neutrino interactions: reconciling the cosmos
  with neutrino oscillations}.
\newblock {\em JCAP}, 1608:067, 2016.

\bibitem{Cherry:2016jol}
John~F. Cherry, Alexander Friedland, and Ian~M. Shoemaker.
\newblock {Short-baseline neutrino oscillations, Planck, and IceCube}.
\newblock 2016.

\bibitem{Forastieri:2017oma}
Francesco Forastieri, Massimiliano Lattanzi, Gianpiero Mangano, Alessandro
  Mirizzi, Paolo Natoli, and Ninetta Saviano.
\newblock {Cosmic microwave background constraints on secret interactions among
  sterile neutrinos}.
\newblock {\em JCAP}, 1707(07):038, 2017.

\bibitem{Chu:2018gxk}
Xiaoyong Chu, Basudeb Dasgupta, Mona Dentler, Joachim Kopp, and Ninetta
  Saviano.
\newblock {Sterile neutrinos with secret interactions?cosmological discord?}
\newblock {\em JCAP}, 1811(11):049, 2018.

\bibitem{Jeong:2018yts}
Yu~Seon Jeong, Sergio Palomares-Ruiz, Mary~Hall Reno, and Ina Sarcevic.
\newblock {Probing secret interactions of eV-scale sterile neutrinos with the
  diffuse supernova neutrino background}.
\newblock {\em JCAP}, 1806:019, 2018.

\bibitem{deGouvea:2019phk}
Andre De~Gouvea, Manibrata Sen, Walter Tangarife, and Yue Zhang.
\newblock {The Dodelson-Widrow Mechanism In the Presence of Self-Interacting
  Neutrinos}.
\newblock 2019.

\bibitem{Babu:1991at}
K.~S. Babu and I.~Z. Rothstein.
\newblock {Relaxing nucleosynthesis bounds on sterile-neutrinos}.
\newblock {\em Phys. Lett.}, B275:112--118, 1992.

\bibitem{Shoemaker:2015qul}
Ian~M. Shoemaker and Kohta Murase.
\newblock {Probing BSM Neutrino Physics with Flavor and Spectral Distortions:
  Prospects for Future High-Energy Neutrino Telescopes}.
\newblock {\em Phys. Rev.}, D93(8):085004, 2016.

\bibitem{Hopkins:2006bw}
Andrew~M. Hopkins and John~F. Beacom.
\newblock {On the normalisation of the cosmic star formation history}.
\newblock {\em Astrophys. J.}, 651:142--154, 2006.

\bibitem{Melnikov:1996iu}
K.~Melnikov and V.~G. Serbo.
\newblock {Processes with the T channel singularity in the physical region:
  Finite beam sizes make cross-sections finite}.
\newblock {\em Nucl. Phys.}, B483:67--82, 1997.
\newblock [Erratum: Nucl. Phys.B662,no.1-2,409(2003)].

\bibitem{Berryman:2018ogk}
Jeffrey~M. Berryman, Andre De~Gouvea, Kevin~J. Kelly, and Yue Zhang.
\newblock {Lepton-Number-Charged Scalars and Neutrino Beamstrahlung}.
\newblock {\em Phys. Rev.}, D97(7):075030, 2018.

\bibitem{deGouvea:2019qaz}
André de~Gouvêa, P.~S.~Bhupal Dev, Bhaskar Dutta, Tathagata Ghosh, Tao Han,
  and Yongchao Zhang.
\newblock {Leptonic Scalars at the LHC}.
\newblock 2019.

\bibitem{Grohs:2020xxd}
E.~Grohs, George~M. Fuller, and Manibrata Sen.
\newblock {Consequences of neutrino self interactions for weak decoupling and
  big bang nucleosynthesis}.
\newblock 2020.

\bibitem{Berezinsky:1975zz}
V.~S. Berezinsky and A.~{\relax Yu}. Smirnov.
\newblock {Cosmic neutrinos of ultra-high energies and detection possibility}.
\newblock {\em Astrophys. Space Sci.}, 32:461--482, 1975.

\bibitem{Capelle:1998zz}
K.~S. Capelle, J.~W. Cronin, G.~Parente, and E.~Zas.
\newblock {On the detection of ultrahigh-energy neutrinos with the Auger
  Observatory}.
\newblock {\em Astropart. Phys.}, 8:321--328, 1998.

\bibitem{Aab:2019auo}
Alexander Aab et~al.
\newblock {Probing the origin of ultra-high-energy cosmic rays with neutrinos
  in the EeV energy range using the Pierre Auger Observatory}.
\newblock {\em JCAP}, 1910(10):022, 2019.

\bibitem{Nelles:2018gqq} 
  A.~Nelles [ARIANNA Collaboration],
\newblock{ARIANNA: Current developments and understanding the ice for neutrino detection}
 \newblock{EPJ Web Conf.}  {\bf 216}, 01008 (2019)
 
 
 \bibitem{Barwick:2016cyg} 
  S.~W.~Barwick,
  \newblock{Progress in the Development of Radio-Cherenkov Neutrino Detectors}
  \newblock{PoS ICRC} {\bf 2015}, 027 (2016).


\bibitem{Murase:2014foa}
Kohta Murase, Yoshiyuki Inoue, and Charles~D. Dermer.
\newblock {Diffuse Neutrino Intensity from the Inner Jets of Active Galactic
  Nuclei: Impacts of External Photon Fields and the Blazar Sequence}.
\newblock {\em Phys. Rev.}, D90(2):023007, 2014.

\end{thebibliography}

\end{document}